\documentclass[conference]{IEEEtran}
\usepackage{floatflt}
\usepackage{amsfonts}
\usepackage{cite}
\usepackage{graphicx}
\usepackage{verbatim}
\usepackage{psfrag}
\usepackage{subfigure}
\usepackage{stfloats}
\usepackage{amsmath}
\usepackage{algorithmic}
\usepackage{dsfont}

\begin{document}

\title{Trading Wireless Information and Power Transfer:\\ Relay Selection to Minimize the Outage Probability}

\author{\IEEEauthorblockN{M. Majid Butt,~Adnan Nasir,~Amr Mohamed and Mohsen Guizani}
\IEEEauthorblockA{Department of Computer Science and Engineering, Qatar University, Doha, Qatar\\
Email: \{majid.butt, adnan.nasir, amrm, mguizani\}@qu.edu.qa}
}

\maketitle

\begin{abstract}
This paper studies the outage probability minimization problem for a multiple relay network with energy harvesting constraints. The relays are hybrid nodes used for \emph{simultaneous} wireless information and power transfer from the source radio frequency (RF) signals. There is a tradeoff associated with the amount of time a relay node is used for energy and information transfer. Large intervals of information transfer implies little time for energy harvesting from RF signals and thus, high probability of outage events. We propose relay selection schemes for a cooperative system with a fixed number of RF powered relays. We address both causal and non-causal channel state information cases at the relay--destination link and evaluate the tradeoff associated with information/power transfer in the context of minimization of outage probability.
\end{abstract}

\section{Introduction}
Wireless sensor networks (WSNs) are generally deployed with inherent energy constraints. They are desired to stay operational for a maximum amount of time. Energy harvesting (EH) devices suit WSN applications because they rely on external charging mechanisms such as solar, wind and RF signals in order to remain active in the network \cite{Medepally_IEEEToWC:2010, Venkata_IEEECS:2010}.

Recent investigations confirm that the use of EH devices improves the performance of the wireless networks. However, the design of sophisticated protocols for EH networks is very critical. In \cite{Yaming}, the authors investigate the optimal relay selection for EH systems using branch and bound algorithm for non-causal case and dynamic programming for the causal case. The relay selection is made on the basis of relay's harvested energy and the largest relative throughput gain.

It has been investigated that information decoding and energy harvesting can be performed from the RF signals and a storage system allows the use of the stored energy for future communication \cite{Gurakan_IEEEToC:2013, Ahmed_IEEEWC:2014}. This concept is termed as simultaneous wireless information and power transfer (SWIPT) where information decoding and energy harvesting is performed by time sharing and/or power splitting \cite{Zhang_IEEEWCOM:2013,Zhang_IEEECOM:2013,Zhou_IEEECOMM:2013}. An amplify-and-forward (AF) wireless cooperative network is
considered in \cite{Ali_IEEEWS:2013} where the relay nodes harvest energy employing one of the time sharing and power splitting protocols. SWIPT framework is further extended to the performance analysis of a large scale network using random geometry approach in \cite{Krikidis_TCOM:2014}.

We explore relay selection schemes based on the available channel state information (CSI) and SWIPT concept. In our framework, we assume that CSI is not available at the relay node on the source-relay link. When CSI is not available at the relay-destination link as well, the relay selection is made on the base of largest available stored energy similar to \cite{Yaming}. Contrary, when CSI is available at the relay-destination link, the relay selection depends both on the amount of stored energy in relays and the channel condition for the relay-destination links. There is a tradeoff between the number of relay nodes involved in information transfer in current time slot and the amount of harvested energy for future use. We evaluate outage performance of both schemes numerically and determine different tradeoffs associated with the number of relays in the network and the energy harvesting efficiency.

The rest of the paper is structured as follows. Section \ref{sect:system model} describes the system model and problem settings for the work. The proposed relay selection schemes are discussed in Section \ref{sect:schemes}. Section \ref{sect:Results} evaluates the performance of the schemes numerically and we conclude with the summary of main results in Section \ref{sect:conclusions}.

\section{System Model and Problem Formulation}
\label{sect:system model}
\begin{figure}
\centering
  	\includegraphics[width=2.9in]{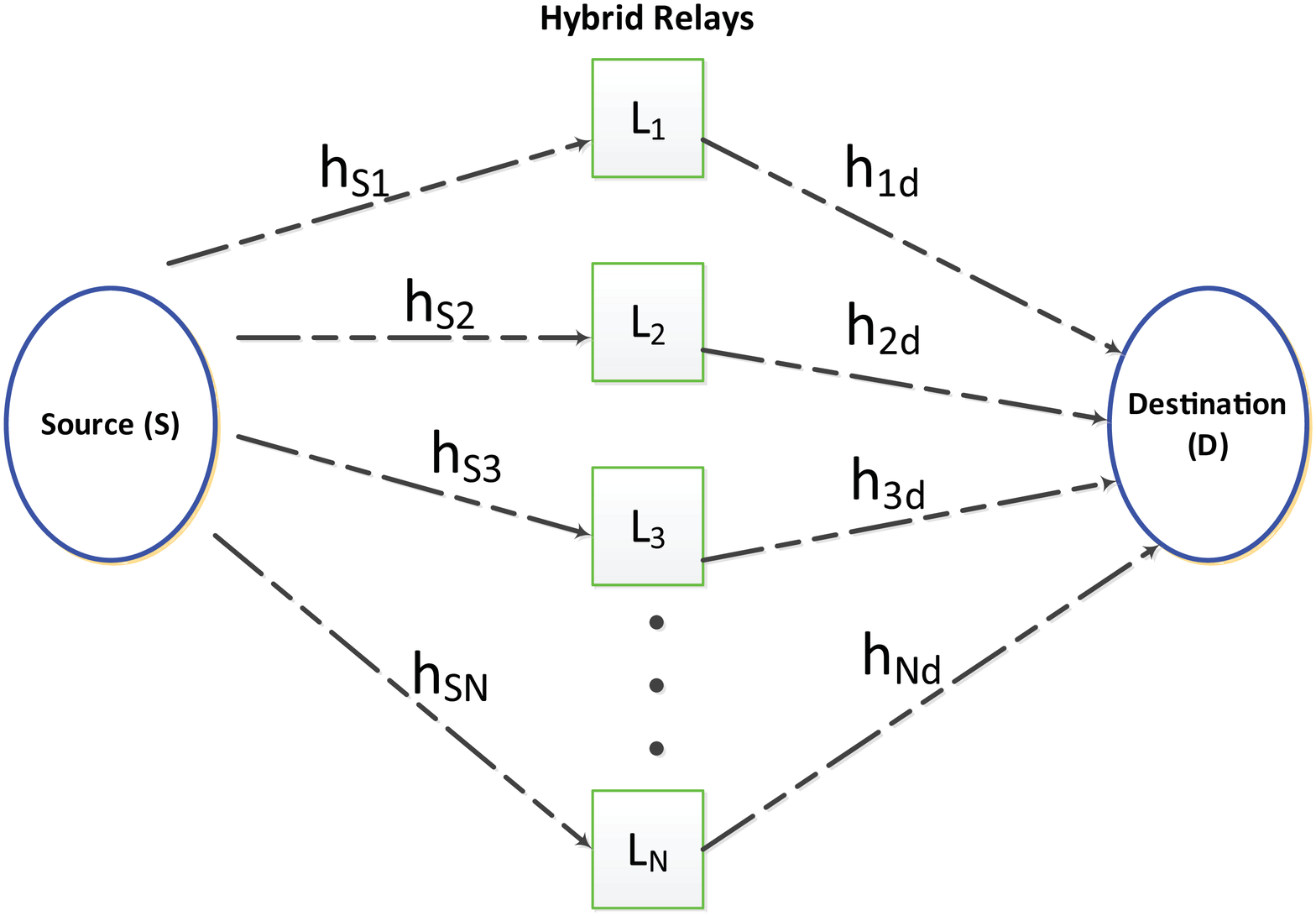}
  \vspace{-0.2cm}
   \caption{Energy cooperation system model.}
   \vspace{-0.4cm}
	\label{fig:system}
\end{figure}

We consider a Decode-and-Forward (DF) strategy based relaying communication system where a source node S communicates with a destination node D in the presence of $N$ relays, represented by symbol $L$ as shown in Fig. \ref{fig:system}. The communication from source to relay and relay to destination takes place in two orthogonal time slots where duration of each slot is denoted by $T$. We assume a fixed transmit power $P_s$ at the source and a broadcast channel for the source-relay communication phase is considered.

The relay nodes are hybrid, i.e., they have the ability to harvest energy as well as retrieve the information from the signal, but we assume (for simplicity) that only one function can be performed in a given time slot $t$ and therefore, no time sharing or power splitting is performed. The hybrid relays include an EH circuit which harvests energy from the transmitted RF signals. The relay selected at time $t$ to forward information to the destination is not available to harvest energy or forward information from the source at time slot $t+1$ due to assumption of orthogonal communication on $S\to L$ and $L\to D$ links. However, the relays other than the selected one are free to receive data, thereby mimicking a full-duplex relaying system \cite{Ikhlef_TVT:2012}. The harvested energy is stored in a battery of an infinite capacity and the energy stored in the battery is assumed to increase and decrease linearly.

We assume independently and identically (iid) distributed fading channels at $S\to L$ and $L\to D$ links which follow block fading model. The received signal $ y_i(t)$ at the relay node $L_i$ is expressed as:
\begin{equation}
y_i(t)= \frac{1}{\sqrt{d_i^2}}\sqrt{P_s}h_{si} x(t)+n(t)
\label{3}
\end{equation}
where $ x(t) $ and $d_i$ denote the normalized information signal from the source and the distance between the transmitter and relay $i$, respectively. $n(t)\sim Z(0,\sigma^2)$ is the Gaussian noise with zero mean and variance $\sigma^2$ while the channel gain coefficient for $S\to L_i$ link is represented by $h_{si}$.

The rate $R_{si}(t)$ provided by $S\to L_i$ link in a time slot $t$ is given by
\begin{equation}
R_{si}(t)= \frac{1}{2}\log_2 \big(1+|h_{si}|^{2}\frac{P_s}{\sigma^2}\big)~.
\label{eqn:rate_equation}
\end{equation}
For a DF relaying strategy, the outage probability that a rate $R$ is not supported by the system is given by
\begin{eqnarray}
P_{\rm out}&=&\mathds{P}\big(\min(R_{si^*},R_{i^*d})<R\big)\\
&=&\mathds{P} \big\{\min\big(\frac{1}{2}\log_2 (1+ | h_{si^*}|^{2}\frac{P_s}{\sigma^2}),\nonumber\\
&&\frac{1}{2}\log_2 (1+ | h_{i^*d}|^{2}\frac{P_{r}}{\sigma^2})\big)<R \big\}\nonumber
\end{eqnarray}
where $L_{i^*}$ denotes the selected relay node. $P_{r}$ is the relay transmit power and $h_{i^*d}$ denotes the channel gain coefficient for the $L_{i^*}\to D$ link.

\subsection{Energy Harvesting Model}
We assume that initially all the relays have sufficient power to retrieve the information and transfer it to the destination. The stored energy will be sufficient to perform the relaying task without the need to harvest energy for the first a few iterations. However, after some iterations the nodes start to deplete their reserved energy and result in an outage or network failure. In order to extend the network lifetime, the energy from the RF source can be intelligently harvested to increase the energy reserves of the relay nodes.

The energy harvested by the $i^{\rm th}$ relay node during a single time slot is given by \cite{Zhou_IEEECOMM:2013, Ali_IEEEWS:2013}
\begin{eqnarray}
E_i^h=\frac{\eta P_s\left | h_{si} \right |^2}{d_i^2}T
\label{4}
\end{eqnarray}
where $ 0< \eta \leq 1 $ is the energy conversion efficiency which depends on the receiver circuit hardware and antenna sensitivity.

\subsection{Problem Settings}
\label{sect:problem_settings}
We express the relay selection problem with the goal to minimize the outage probability. Practically, there is some hardware dependent minimum signal energy threshold that governs whether the node can sense the signal (and harvest energy) or not. If the signal received is less then the threshold, the relays cannot perform energy harvesting. Without loss of generality and to concentrate on the system design, we set it equal to zero.

We formulate the outage probability minimization problem for a multiple relay network:
\begin{eqnarray}
&\min&~P_{\rm out}\\
&{\rm s.t}.&~
\begin{cases}
N= c, \quad c\in \mathbb{N}\\
R\geq 0
\end{cases}
\end{eqnarray}
$c$ is a constant representing a fixed number of relays in the system. As network is operated by energy harvested from the source RF signals, there must be sufficient (or at least one) charged relay nodes in a given time slot to be able to forward the signal successfully in order to avoid the outage event. Thus, there is a tradeoff between the number of relay nodes in EH mode and the number of nodes available for information transfer for a given transmission. The larger the number of nodes in EH mode, the more inefficient is the use of $L\to D$ link for information transfer in the current time slot, but more energy is available for information transfer in future. This is the main reason that EH communication focuses on meeting the neutrality constraint\footnote{Neutrality constraint refers to the goal of using the resources in such a way that the probability of availability of resources for future use is maximized.} in contrast to making the best use of available resources using opportunistic communications solely in the current time slot.


\section{Relay Selection Schemes}
\label{sect:schemes}
We assume that CSI is not available at $S \to L$ link. Regarding CSI on $L \to D$ link, we consider two cases which govern the relay selection strategy:
\begin{itemize}
\item The CSI at relay is causal and not available.
\item The CSI is known before transmission at $L\to D$ link.
\end{itemize}
\vspace{-0.2cm}
\subsection{Single Relay Selection (SRS)}
First, we assume that CSI is not available at the relay for transmission to the destination and therefore, the selected relay transmits with a fixed power $P_r$. The forwarding relay is selected solely based on the stored energy at the relay nodes.
In this case, only a single relay node $L_{i^*}$ is selected out of $N$ nodes to decode and forward the information. We take this case as a baseline and compare results with our scheme in the next section.
Similar to \cite{Yaming}, the node $L_{i^*}$ with the maximum stored energy from $N$ candidate nodes is selected such that:
\begin{equation}
i^*=\arg\max_{i} \big(E_i^{\rm store}(t) - E_r\big)^+
\label{eqn:scen1}
\end{equation}
where $E_r$ is the energy spent due to transmission with fixed power $P_r$ and $E_i^{\rm store}(t)$ denotes the stored energy for the relay $L_i$ at time $t$. Note that the relay selection is performed before the signal reception from the source and therefore, all other relays can harvest energy from the received RF signal using harvesting circuit. If $E_i^{\rm store}(t) < E_r,\forall i$, no node is selected and all $N$ nodes harvest energy. For the case, $R_{si^*}<R$, node $i^*$ is unable to decode information from the source and results in an outage without making a transmission on $L\to D$ link.

All the nodes except $L_{i^*}$ harvest energy depending on the received signal strength from the source such that
\begin{equation}
E_j^{\rm store}(t+1) = E_j^h(t)+ E_j^{\rm store}(t), \quad j\ne i^*~.
\end{equation}
As mentioned in Section \ref{sect:system model}, the selected node $i^*$ is not a candidate for selection in time slot $t+1$ for both proposed schemes and therefore, the energy update is only meaningful for time slot $t+2$. Note that $E_{i^*}^{\rm store}(t+1)=E_{i^*}^{\rm store}(t)$.

Thus, the corresponding stored energy for node $i^*$ is given by
\begin{equation}
E_{i^*}^{\rm store}(t+2)= E_{i^*}^{\rm store}(t+1)-E_{r}~.
\end{equation}
If we increase $N$, we have more relays to choose $i^*$ for data transfer to the destination. This results in decrease in outage.

\subsection{Multiple Relay Selection (MRS)}
\label{sect:MRS}
In this case, we assume that CSI is known at the relay node for transmission to the destination. However, signal from the source is received in time slot $t$ and transmitted to the destination in time slot $t+1$. Based on the available information, we propose a 2-step relay selection policy.

In the first step, a subset $\Gamma$ of $M$ relays is selected out of $N$ relays such that
\begin{equation}
\Gamma^{M\times 1}=\{i:E_i^{\rm store}\geq \gamma_M\}
\label{eqn:csi_gamma}
\end{equation}
where $\gamma_M$ defines the stored energy of the node with $M^{th}$ largest stored energy.
Equation $(\ref{eqn:csi_gamma})$ states that $\Gamma$ contains elements with $M$ largest stored energies out of $N$ relays.
As fading distribution is i.i.d and CSI for the next time slot is not available at time $t$, the selection is based on the known stored battery condition for the relays. All the nodes $i\in \Gamma$ (attempt to) decode the information from the source and cannot harvest energy in time slot $t$ while rest of the $N-M$ nodes harvest energy. We limit the cardinality of the set $\Gamma$ to a fixed value $M\leq N$ where $M$ is a system parameter to be optimized.

Then, a set $\Lambda$ is selected out of $M$ nodes that can retrieve the information from the signal on $S\to L$ link such that
\begin{equation}
\Lambda=\{i:i\in \Gamma,R_{si}>R\}
\label{eqn:lambda}
\end{equation}
As CSI at relay nodes in $\Lambda$ is available at the time of transmission in time slot $t+1$, a single relay $L_{i^*}$ from the set $\Lambda$ is selected such that
\begin{equation}
{i^*}=\arg\max_{i\in \Lambda} \big(E_i^{\rm store}(t+1) - E_r^i(t+1)\big)^+
\label{eqn:scen2_selection}
\end{equation}
where $E_r^{i}(t+1)$ results from $P_r^{i}(t+1)$ and given by
\begin{equation}
P_{r}^{i} = \frac{(2^{2R}-1)\sigma^2}{|h_{id}|^{2}}
\label{eqn:power}
\end{equation}
If $E_i^{\rm store}(t+1) < E_r^i(t+1),\forall i\in \Lambda$, no node is selected for transmission which results in outage, but avoids energy loss due to unsuccessful transmission from node $i^*$.

If $E_{i^*}^{\rm store}(t+1) > E_r^{i^*}(t+1)$, the stored energy for node $L_{i^*}$ is updated such that
\begin{eqnarray}
E_{i^*}^{\rm store}(t+2)= E_{i*}^{\rm store}(t+1)-E_r^{i^*}(t+1)
\end{eqnarray}
The rest of the nodes harvest and store energy depending on the received signal strength from the source such that
\begin{eqnarray}
E_j^{\rm store}(t+1)=
\begin{cases}
E_j^h(t)+ E_j^{\rm store}(t),&   j \notin \Gamma\\
E_j^{\rm store}(t),&   j \in \Gamma,j \ne i^* ~.
\end{cases}
\end{eqnarray}
We notice that parameter $M$ controls the outage probability for a fixed $N$. There is a tradeoff associated with selection of $M$. Increasing $M$ makes the relay selection in (\ref{eqn:scen2_selection}) more opportunistic due to large cardinality of set $\Lambda$ and more freedom in choosing $i^*$ in (\ref{eqn:scen2_selection}). However, note that all $i\in \Gamma$ do not harvest energy and their storage level remains the same. In this work, we assume no leakage factor but practically, there is a leakage in storage for every node in each time slot even if the node is not transmitting. Large $M$ implies that less number of nodes are charging their batteries and therefore, the storage at system level keeps on decreasing and causes more outage (network failure). Therefore, there is an optimal $M\leq N$ for the proposed scheme which maximizes the performance.

Given that we have MRS policy $\pi(M,N)$ for relay selection, the parameter optimization problem is formulated by
\begin{eqnarray}
M^*(R,\eta) &=& \arg\min_{\pi(M,N), 0<M\leq N} P_{\rm out} \\
&{\rm s.t.} &
\begin{cases}
N=c,& c\in \mathbb{N}\\
R\geq 0
\end{cases}
\end{eqnarray}
The value of $M^*$ depends on the number of relays in the system $N$, energy harvesting efficiency $\eta$ and rate region. At small $R$, very large $M$ (and set $\Lambda$) is not helpful to achieve multiuser diversity and factor $\eta$ dominates the outage performance behaviour. However, large $M$ improves the performance at large $R$ as evaluated numerically in Sec. \ref{sect:Results}.

\section{Numerical Results}
\label{sect:Results}
We assume independent Rayleigh fading channels with mean 1 on both $S\to L$ and $L\to D$ channels. 20000 iterations are performed to compute outage probability numerically for the simulation results. The relays are assumed to be equidistant from the source with $d$ equals one. $P_s$ is fixed to 10 dbW.

\begin{figure}
\centering
  	\includegraphics[width=3.2in]{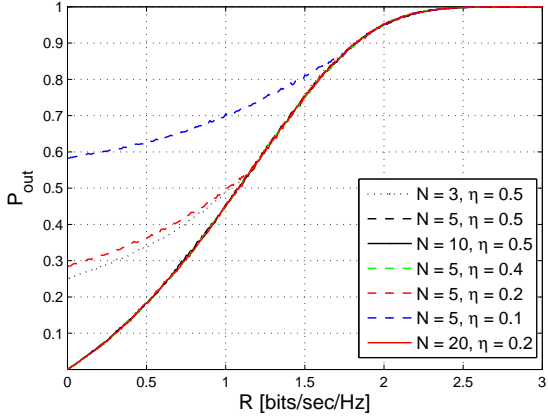}
  \vspace{-0.3cm}
   \caption{Outage probability for SRS scheme for different $N$ and $\eta$ values.}
   \vspace{-0.4cm}
	\label{fig:SRS}
\end{figure}

Fig. \ref{fig:SRS} shows the outage probability for the SRS scheme when $N$ is fixed. As CSI is not available at the relay node, $P_r$ is fixed to 10 dBW. As expected, the outage probability increases as $R$ increases. The number of relays $N$ and energy harvesting efficiency factor $\eta$ are important factors to characterize the scheme. For a fixed value of $N$, a decrease in $\eta$ results in decreased harvested energy for the relay nodes. When $\eta$ is decreased initially, $P_{\rm out}$ remains the same as for $N=5$ case with $\eta=0.5$ and $\eta=0.4$, which implies that at least a single node is always available with enough harvested energy. However, if $\eta$ is too small, the probability that no relay has enough energy to make a successful transmission increases as evident for the case $N=5, \eta=0.2$ in Fig. \ref{fig:SRS}. When $\eta$ is very small, the outage is observed even for very small $R$ as the selected relay must have enough energy to transmit with power $P_r=10$ dBW regardless of $R$. This region can be termed as power limited region where the outage performance is dominated by the harvested energy as compared to the large rate region where the channel distribution determines the outage behaviour and power limitation effect almost vanishes for different $\eta$.

The same effect is observed with small $N$ where the effect of small $\eta$ is even more pronounced as $N-1$ relays harvest energy in a single time slot. Limiting $N$ exaggerates the power limitation effect due to poor energy harvesting efficiency. As $\eta$ in practically available systems is too low, it is important to have large $N$ to reduce the effect of small $\eta$. For example, $N=20$ in Fig.~\ref{fig:SRS} improves outage performance considerably at small $R$ as compared to $N=5$ case when $\eta=0.02$.

\begin{figure}
\centering
  	\includegraphics[width=3.2in]{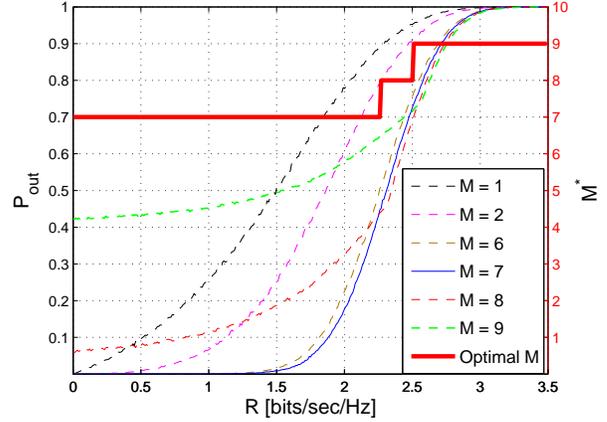}
  \vspace{-0.3cm}
   \caption{Outage probability for MRS scheme for parameters $N=10$, $\eta=0.05$ and different $M$. The optimal $M$ (with y-axis on the right side) curve has been plotted as well, which shows the value of optimal $M$ for every $R$.}
   \vspace{-0.5cm}
	\label{fig:MRS}
\end{figure}

Fig. \ref{fig:MRS} shows the outage performance for MRS case. As CSI is known at the relay node, $P_r^{i}$ is determined by (\ref{eqn:power}). For a fixed $N$, we plot the outage probability curves for different values of $M$ and determine the optimal value $M^*$ numerically. As discussed in Section \ref{sect:MRS}, the outage probability for $M>M^*$ is not optimal due to sub-optimality in energy harvesting from RF signals while $M<M^*$ results in too small group of candidate relays to exploit multiuser diversity. For the numerical example with $N=10$, $M=7$ provides the optimal outage performance at a small rate. Though, $M=8, M=9$ perform marginally better than $M=7$ at large $R$, the incremental gain is so small that $M=7$ can be approximated as the optimal $M$ value for all $R$. Fig. \ref{fig:MRS} shows the optimal $M$ curve (with $M^*$ on the right side y-axis) in different rate regions where $M^*=7$ up to $R=2.25$, then $M=8$ becomes optimal while $M=9$ is the optimal solution for $R>2.55$.

\begin{figure}
\centering
  	\includegraphics[width=3.2in]{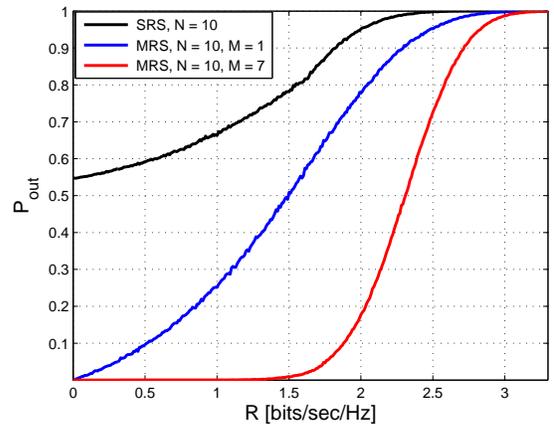}
  \vspace{-0.3cm}
   \caption{Outage comparison of SRS and MRS schemes for a fixed $N=10$ and $\eta=0.05$.}
   \vspace{-0.6cm}
	\label{fig:SRS_MRS}
\end{figure}

Fig.~\ref{fig:SRS_MRS} compares SRS and MRS schemes for the same value of $N$. MRS outperforms SRS scheme even for $M=1$ case thanks to power allocation according to available CSI at $L-D$ link. However, performance improves considerably when $M=M^*$ for the MRS scheme.

\vspace{-0.1cm}
\section{Conclusions}
\label{sect:conclusions}
This work investigates the relay selection problem for energy harvesting communication system. The relay nodes are dual nodes with energy harvesting and wireless information transfer capabilities. Based on the channel state information availability, two simple relay selection schemes are discussed. The outage performance of both schemes is investigated numerically. The results show that the availability of channel state information at relays improves performance considerably. Moreover, energy harvesting efficiency of the relay nodes is a limiting factor for the outage performance of the schemes and there is a tradeoff associated involving number of relays in the system versus energy harvesting efficiency of the relays. As an extension to this work, we will focus on studying analytical models of the schemes and evaluate the performance as a function of energy harvesting efficiency and network size.
\vspace{-0.1cm}
\section*{Acknowledgement}
This publication was made possible by NPRP 5-782-2-322 from the Qatar National Research Fund (a member of The Qatar Foundation). The statements made herein are solely the responsibility of the authors.

\bibliographystyle{IEEEtran}
\bibliography{bibliography}

\begin{thebibliography}{10}
\providecommand{\url}[1]{#1}
\csname url@samestyle\endcsname
\providecommand{\newblock}{\relax}
\providecommand{\bibinfo}[2]{#2}
\providecommand{\BIBentrySTDinterwordspacing}{\spaceskip=0pt\relax}
\providecommand{\BIBentryALTinterwordstretchfactor}{4}
\providecommand{\BIBentryALTinterwordspacing}{\spaceskip=\fontdimen2\font plus
\BIBentryALTinterwordstretchfactor\fontdimen3\font minus
  \fontdimen4\font\relax}
\providecommand{\BIBforeignlanguage}[2]{{%
\expandafter\ifx\csname l@#1\endcsname\relax
\typeout{** WARNING: IEEEtran.bst: No hyphenation pattern has been}%
\typeout{** loaded for the language `#1'. Using the pattern for}%
\typeout{** the default language instead.}%
\else
\language=\csname l@#1\endcsname
\fi
#2}}
\providecommand{\BIBdecl}{\relax}
\BIBdecl

\bibitem{Medepally_IEEEToWC:2010}
B.~Medepally and B.~Mehta, N., ``Voluntary energy harvesting relays and
  selection in cooperative wireless networks,'' \emph{IEEE Transactions on
  Communications}, vol.~9, no.~11, pp. 3543--3553, 2010.

\bibitem{Venkata_IEEECS:2010}
P.~Venkata, S.~Nambi, R.~Prasad, and I.~Niemegeers, ``Bond graph modeling for
  energy-harvesting wireless sensor networks,'' \emph{IEEE Transactions on
  Communications}, vol.~45, no.~9, pp. 31--38, 2012.

\bibitem{Yaming}
Y.~Luo, J.~Zhang, and K.~B. Letaief, ``Relay selection for energy harvesting
  cooperative communication systems,'' in \emph{IEEE Global communications
  conference (Globecom)}, Atlanta, GA, USA, Dec. 2013.

\bibitem{Gurakan_IEEEToC:2013}
B.~Gurakan, O.~Ozel, Y.~Jing, and S.~Ulukus, ``Energy cooperation in energy
  harvesting communications,'' \emph{IEEE Transactions on Communications},
  vol.~61, no.~12, pp. 4884--4898, 2013.

\bibitem{Ahmed_IEEEWC:2014}
I.~Ahmed, A.~Ikhlef, R.~Schober, and R.~K. Mallik, ``Power allocation for
  conventional and buffer-aided link adaptive relaying systems with energy
  harvesting nodes,'' \emph{IEEE Transactions on Wireless Communications},
  vol.~13, no.~3, pp. 1182--1195, March 2014.

\bibitem{Zhang_IEEEWCOM:2013}
R.~Zhang and C.~K. Ho, ``{MIMO} broadcasting for simultaneous wireless
  information and power transfer,'' \emph{IEEE trans. Wireless Communications},
  vol.~12, no.~5, pp. 1989--2001, May 2013.

\bibitem{Zhang_IEEECOM:2013}
L.~Liu, R.~Zhang, and K.-C. Chua, ``Wireless information and power transfer: A
  dynamic power splitting approach,'' \emph{IEEE trans. Communications},
  vol.~61, no.~9, pp. 3990--4001, Sep. 2013.

\bibitem{Zhou_IEEECOMM:2013}
X.~Zhou, R.~Zhang, and C.~K. Ho, ``Wireless information and power transfer:
  Architecture design and rate-energy tradeoff,'' \emph{IEEE Transactions on,
  Communications}, vol.~61, no.~11, pp. 4754--4767, 2013.

\bibitem{Ali_IEEEWS:2013}
N.~Ali, A., Z.~Xiangyun, D.~Salman, and K.~Rodney, A., ``Relaying protocols for
  wireless energy harvesting and information processings,'' \emph{IEEE
  Transactions on Wireless Communications}, vol.~12, no.~7, pp. 3622--3636,
  2013.

\bibitem{Krikidis_TCOM:2014}
I.~Krikidis, ``Simultaneous information and energy transfer in large-scale
  networks with/without relaying,'' \emph{IEEE trans. Communications}, vol.~62,
  no.~3, pp. 900--912, March 2014.

\bibitem{Ikhlef_TVT:2012}
A.~Ikhlef, J.~Kim, and R.~Schober, ``Mimicking full-duplex relaying using
  half-duplex relays with buffers,'' \emph{IEEE Transactions on Vehicular
  Tech.}, vol.~61, no.~7, pp. 3025--3037, September 2012.

\end{thebibliography}


\end{document}